\documentstyle[11pt,newpasp,twoside,epsf]{article}

\reversemarginpar

\begin{document}

\title{The Most Detailed Picture Yet of an Embedded High-mass YSO}

\author{L. J. Greenhill, M. J. Reid}
\affil{Harvard-Smithsonian Center for Astrophysics, 60 Garden St,
Cambridge, MA 02138, USA}

\author{C. J. Chandler}
\affil{NRAO, P.O. Box O, Socorro, NM 87801, USA}

\author{P. J. Diamond}
\affil{MERLIN/VLBI National Facility, Jodrell Bank Observatory, Macclesfield, SK11 9DL, UK}

\author{M. Elitzur}
\affil{Department of Physics \& Astronomy, University of Kentucky, Lexington, KY 40506, USA}

\begin{abstract}
High-mass star formation is not well understood chiefly because examples are deeply
embedded, relatively distant, and crowded with sources of emission. Using VLA and VLBA
observations of H$_2$O and SiO maser emission, we have mapped in detail the structure and
proper motion of material 20--500\,AU from the closest high-mass YSO, radio source~{\sf I} in
the Orion KL region.  We observe streams of material driven in a rotating, wide angle, bipolar wind from
the surface of an edge-on accretion disk.  The example of source~{\sf I} provides strong
evidence that high-mass star formation proceeds via accretion.
\end{abstract}

\section{Introduction}

High-mass star formation is poorly understood because theory has yet to
resolve the balance of radiation pressure, gravity, magnetic energy, and thermal heating
(e.g., McKee \& Tan 2002). Moreover, examples amenable to detailed study but 
young enough to have not yet formed HII regions are rare.  
First, massive YSOs are deeply embedded, form in clusters, and heat large
amounts of  surrounding dust and gas.  As a result, the signatures of disks and outflows at
small radii are easily lost in ``background clutter.''  Second, regions of high-mass star
formation are sufficiently distant ($\ga 500$\,pc) that infrared and radio observations of
thermal gas and dust are often confusion-limited.  Third, massive young stellar objects
evolve rapidly, and by the time the surrounding medium is dispersed, accretion disk and
outflow structures have been at least partly dispersed.   In two of the best understood cases,
G192.16$-$3.82 (Shepherd \& Kurtz 1999) and IRAS\,20126+4104 (Zhang et al.\ 1998; Cesaroni et
al.\ 1999), disk-like structures have been detected, but the rotation curves 
are barely resolved and the emission arises at large enough 
radii ($>1000$\,AU) that in depth study is difficult.

The infrared Kleinemann-Low (KL) nebula in Orion is the 
%first (Kleinmann \& Low 1967),
nearest ($\sim 500$\,pc), and most heavily studied region of high-mass star formation.  The
KL region is crowded, containing 16 identified mid-infrared peaks across $\sim 10^4$\,AU
(Gezari, Backman, \& Werner 1998). Offset $\sim 0\rlap{.}''5$ south of the prominent 
peak, IRc2, is radio source~{\sf I}, first detected by Churchwell et al.\ (1987). It has
no  infrared counterpart, but it does power compact distributions of  
SiO and H$_2$O masers, and it has been identified as a probable deeply embedded,
massive protostar or YSO (Gezari 1992; Menten \& Reid 1995).

The strongest SiO maser emission arises from the v=1 vibrational state.
Because of excitation requirements ($T >10^3$\,K, $n({\rm H_2})
\sim 10^{10\pm1}$\,cm$^{-3}$), it must originate close to the YSO.
The first VLBI maps revealed an {\sf X}-shaped SiO emission locus
extending outward 20--70\,AU from source~{\sf I}. The {\sf X} also lay
at the center of a $200\times600$\,AU expanding patch of H$_2$O masers
distributed in two lobes each comprising red and blueshifted emission
and bracketing source~{\sf I} (Greenhill et al.\ 1998; see also Gaume et
al.\ 1998; Doeleman, Lonsdale, \& Pelkey 1999).

Taken alone, the distribution of SiO emission could have been interpretted equivalently in
two ways. First, it could represent the turbulent limb of a high-velocity biconical outflow
with a southeast-northwest axis.  Second, it could represent hot material close to the top
and bottom surfaces of an edge-on disk with a northeast-southwest rotation axis.  Greenhill
et al. and Doeleman et al. adopted the biconical outflow model, in part because the
high-velocity outflow observed in the KL region on scales of $10^4$\,AU (e.g., Allen \&
Burton 1993; Schultz et al. 1999) subtended the opening angle of the putative SiO maser
cone.  In addition, Greenhill et al. (1998) noted that the distributions of proper motions
and line-of-sight velocities among the surrounding H$_2$O masers were not readily consistent
with the edge-on disk model and proposed instead that the H$_2$O masers lay on the surface
of a nearly edge-on, inflating, equatorial doughnut-like shell.  This shell could arise
from a slow stellar equatorial wind advancing into the surrounding medium or from 
photo-evaporation of an accretion disk.  Because the major axis of this shell was aligned
with the the so-called ``18~km\,s$^{-1}$'' outflow in the KL region (e.g.,
Genzel \& Stutzki 1989), it appeared that source~{\sf I} might drive the two most prominent
outflows in the region.

Despite circumstantial evidence tying source~{\sf I} to large scale outflows in the KL
region, ambiguity has remained.  On the one hand, Menten \& Reid (1995) noted that another
infrared source, {\sf n}, lies closer to the previously estimated center of the flows
(Genzel et al. 1981).  Moreover, high-velocity gas far to the northwest is blueshifted,
while the corresponding outflow cone near source~{\sf I} is redshifted.  On the other hand,
the compact bipolar radio lobes of source~{\sf n} are oriented north-south,
$40-50^\circ$ from the principal axes of the low and high-velocity large scale flows in the 
KL region. 

To better understand the KL region  and to test the biconical outflow model proposed for
source~{\sf I}, we remapped the SiO and H$_2$O maser emission, covering three $J=1\rightarrow
0$ transitions of SiO (v=0, 1, \& 2) and the full velocity range of H$_2$O emission.  (The
original maps presented by Greenhill et al. were significantly limited in dynamic range,
especially at velocities close to the systemic velocity.)  We measured the proper motion of
the SiO v=1 \& 2 emission, observing monthly over $\sim 3.5$ years, which is $\sim 30\%$ of the
dynamical crossing time of the outflow.  Here we discuss proper motions over a four month
interval.

\section{Observations and Data Reduction}

We observed the v=1 \& 2 $J=1\rightarrow 0$ transition of SiO with the VLBA,
and the v=0 $J=1\rightarrow 0$ transition of SiO and $6_{16}\rightarrow 5_{23}$ transition
of H$_2$O with the VLA in its largest configuration and including the nearby Pie Town
VLBA antenna.  We obtained angular resolutions of $\sim 0.2$\,milliarcseconds
(mas) with the VLBA and $\sim 50$\,mas with the VLA.  The spectral channel spacings in the
VLBA and VLA imaging were $\sim 0.4$\,km\,s$^{-1}$ and
$\sim 2.6$\,km\,s$^{-1}$, respectively.

Using the VLBA, we observed the SiO v=1 \& 2 emission (from $V_{\rm
LSR} \sim -15$ to +25\,km\,s$^{-1}$) simultaneously to enable registration of both lines 
to $\sim 0.1$\,mas.
Because emission from source~{\sf I} itself is thermal, it could
not be detected with the  VLBA.  To locate the position of the YSO on the
VLBA maps, we convolved the maps with a circular beam comparable in size
to the beam of the VLA at $\lambda 7$mm and overlayed the degraded VLBA
map on the VLA map of Menten \& Reid (1995), who detected and registered
the SiO v=1 and thermal continuum emission of the YSO.  The uncertainty
in this registration is $\sim 10$\,mas.

%The RMS noise in our
%images was $\sim 10$ to 20\,mJy and the dynamic range limited by sampling of the {\it
%(u,v)}-plane to a few $\times 10^3$. (The peak maser flux density was $\sim 600$\,Jy.)  

Using our VLA observations, we measured the positions of the SiO v=0 masers ($\sim -10$
to +22\,km\,s$^{-1}$) with respect to the v=1 emission, which was observed simultaneously
so as to provide an astrometric reference.  The uncertainty in position relative to
source~{\sf I} is $\sim 10$\,mas.  The positions of the H$_2$O masers ($\sim -10$ to
+16\,km\,s$^{-1}$) were measured with respect to a nearby quasar and compared to the absolute
position of source~{\sf I} (Menten \& Reid 1995).  The uncertainty in this comparison is
30\,mas.

\section{Results and Discussion}

Our more extensive mapping of the SiO and H$_2$O maser emission
now supports the edge-on disk model previously discarded.  We
propose that the v=1 \& 2 SiO maser emission traces material streaming in a
rotating funnel-like flow from the upper and lower surfaces of an accretion disk of a massive YSO. 
The v=0 SiO and H$_2$O masers lie in a bipolar outflow along the disk rotation axis.  

\subsection{$R<70$\,AU}

In the most recent observations, the v=1 \& 2 SiO emission traces an {\sf X} as before
(Figure\,1), but we note two important new details.  First, there is a ``bridge'' of maser
emission extending from the base of the south arm to the base of the west arm, and there is
a gradient in line-of-sight velocity along the bridge. Second, the arms are not radial. 
These
new findings may be consequences of our having now mapped the v=2 as well as v=1 emission.
(For instance, the bridge is outlined principally by v=2 emission.)  Both
findings are difficult to explain in the context of the biconical flow model.  However, they
are readily explained in the context of the edge-on disk model. The bridge and associated
velocity gradient are natural signatures of emission from the front side of a rotating disk
that is tipped down slightly to the southwest.  The canting of the arms, so that they are not radial,
is also suggestive of reflection symmetry about a plane (i.e., disk).

\begin{figure}[!th]
\plottwo{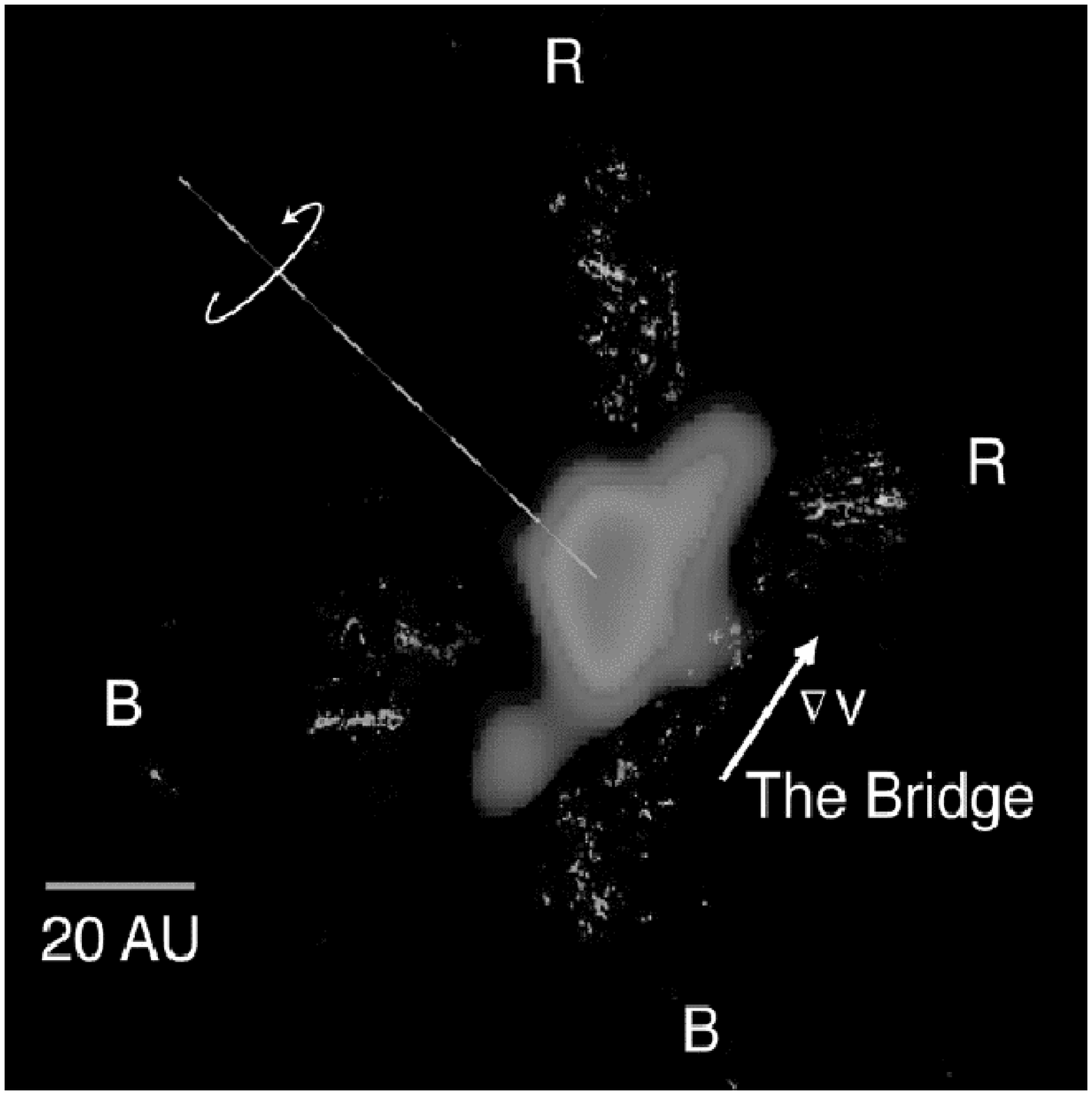}{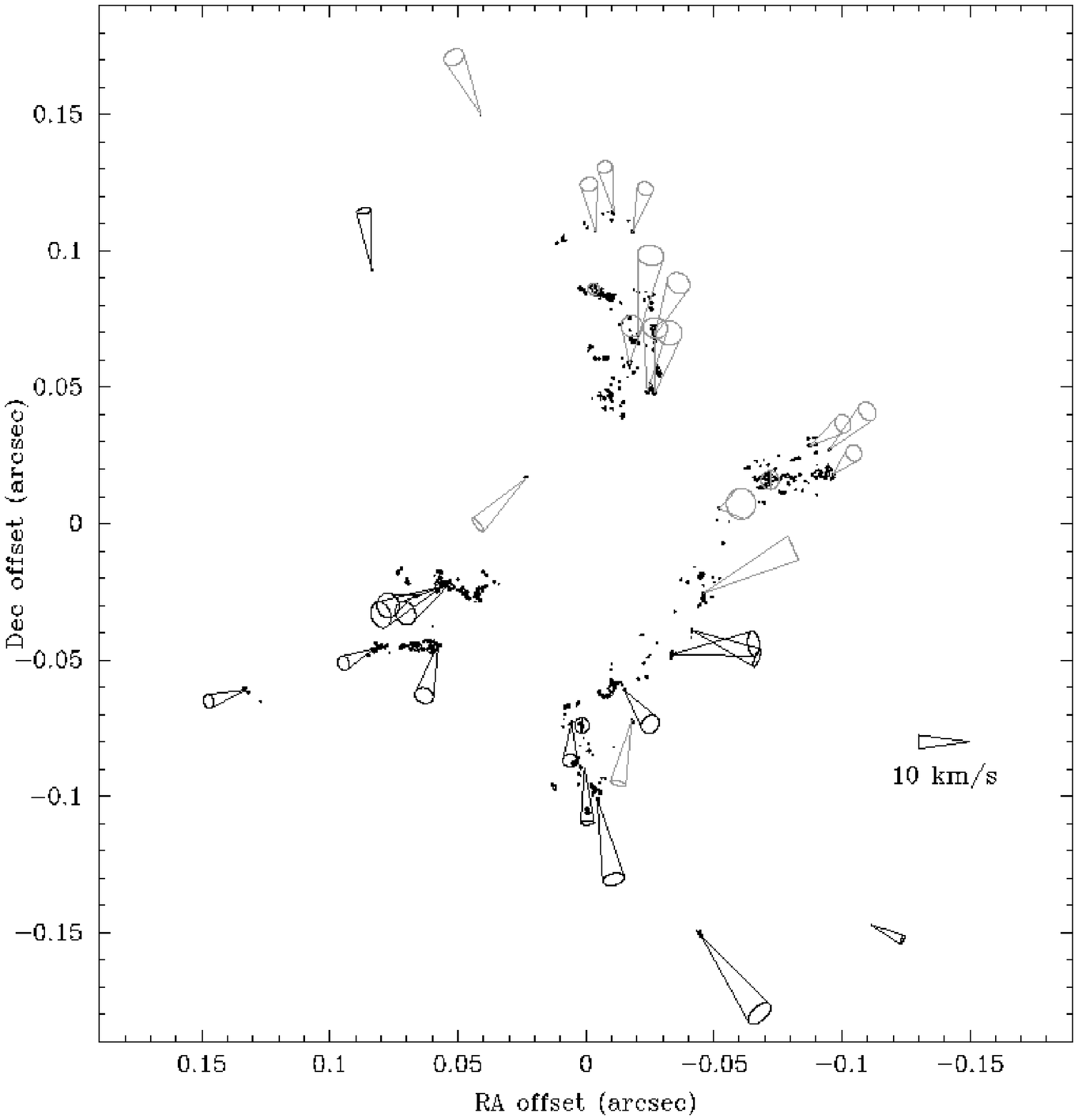}  
\caption{{\it (Left:)} Vibrationally excited SiO masers
forming an {\sf X} and $\lambda 7$mm continuum emission (grayscale) observed by Menten \&
Reid (in prep).   Emission redshifted (R) with respect to the systemic velocity
($V_{\rm LSR}=5$\,km\,s$^{-1}$) lies to the north and west. Blueshifted emission (B)
is opposite. The line emission probably originates in rotating high density
material driven from the surface of an edge-on disk. The continuum  
probably arises at least in part from H$^-$ free-free disk emission. {\it
(Right:)} Proper motions of v=1 \& 2 SiO maser clumps (Chandler et al., in prep). The lengths of the cones
indicate 3-D velocities.  The aspects of the cones indicate inclinations with respect
to the line of sight.  Blueshifted motions are black, redshifted are gray.}
\end {figure}

In order to estimate the proper motions of maser clumps, we identified one in each quadrant
of the {\sf X} that persisted in each epoch and maintained the same line-of-sight velocity. 
We overlayed images for the four epochs and registered them to achieve zero mean proper
motion for the source as a whole.  Most proper motions are 10 to 15\,km\,s$^{-1}$,
though  some clumps are nearly stationary on the sky.  (One km\,s$^{-1}$ over four
months corresponds to $\sim 1$ VLBA beamwidth.)   The maximum observed 3-D velocity is
$\sim 23$\,km\,s$^{-1}$.  The proper motions of maser clumps are chiefly along the
four arms of the {\sf X} (Figure\,1).  In the bridge, the motions are both outward
and tangential, indicative of rotation and strongly in support of the edge-on disk model
(Chandler et al., in prep).   

We suggest that the arms of SiO emission represent the limbs of a bipolar funnel-like 
outflow (Figure\,2), probably the turbulent shocked interface between outflowing and
accreting material.  In a rotating system, strong maser emission is expected along the limbs
because that is where the longest maser gain paths lie owing to projection effects.  
Additional emission from the bridge may mark the nearside of the outflow wall, where at
the base, higher densities or temperatures could compensate for otherwise shorter
gain paths.  The dynamical mass of the YSO is difficult to estimate because the maser
material is not in simple Keplerian rotation.
If the observed 3-D velocity is greater than the
escape velocity, then the enclosed mass is $\ga 6$\,M$_\odot$, for a
radius of 25\,AU and velocity of 25\,km\,s$^{-1}$.
We note that the 3-D motion of the maser material is helical. This may indicate the presence
of a strong magnetic field that is probably anchored to the accretion disk, since massive
stars are radiative and may not generate their own field.  If the magnetic and kinetic
energy densities are of the same order, then the field is on the order of 1\,G, for a gas density
of $10^{10}$\,cm$^{-3}$.

\subsection{$R>70$\,AU}

The v=0 SiO and H$_2$O maser emission arises chiefly from two ``polar caps'' that subtend the
$\sim 90^\circ$ opening angle of the outflow along the disk rotation axis (Figure\,2).  The velocity
structure of the emission is somewhat disorganized and difficult to model in detail.  However, the
two lobes (northeast and southwest) display the same ranges of velocity, indicative of
outflow in the plane of the sky.   Within each lobe, the bulk of the emission displays a
red-blue asymmetry across the rotation axis in the same sense as the  disk rotation close to
source~{\sf I}, although further study is
required.  Otherwise, it may signify coupling between the velocity field downstream in the
outflow and the underlying accretion disk, possibly via magnetic processes.

Substantial overlap of H$_2$O maser and SiO v=0 emission along the line of sight is surprising
because maser action in each species requires quite different densities,
$10^{8-10}$~cm$^{-3}$ vs $10^{5-6}$~cm$^{-3}$, respectively.  Moreover, because 
molecule reformation on dust grains is essential for H$_2$O
maser action behind shocks, gas phase SiO would be depleted.
In principle, the H$_2$O and v=0 SiO
maser volumes could be intermingled if the flow were inhomogeneous and if shock induced grain
sputtering enhanced gas phase concentrations of SiO.  However, the requisite $> 10^2$
density contrast between emitting regions would be difficult to explain in the apparent absence of
high-velocity ($\gg 50$\,km\,s$^{-1}$) gas motions.  

We suggest that the wall of the outflow traced by v=1 \& 2 SiO masers at radii
$<70$\,AU extends to larger radii where it supports H$_2$O maser emission.  The outflow
itself has low enough density to support the v=0 SiO maser emission, which lies along the
same line of sight as the H$_2$O emission in projection.  The ordered velocity structure of
the v=1 \& 2 SiO maser emission and the relatively complicated velocity structure of the 
H$_2$O and v=0 SiO maser emission at larger radii may be a consequence of the increased
effects turbulence or collisions with inflowing or ambient material downstream in the
outflow.

\begin{figure}[!th]
\plottwo{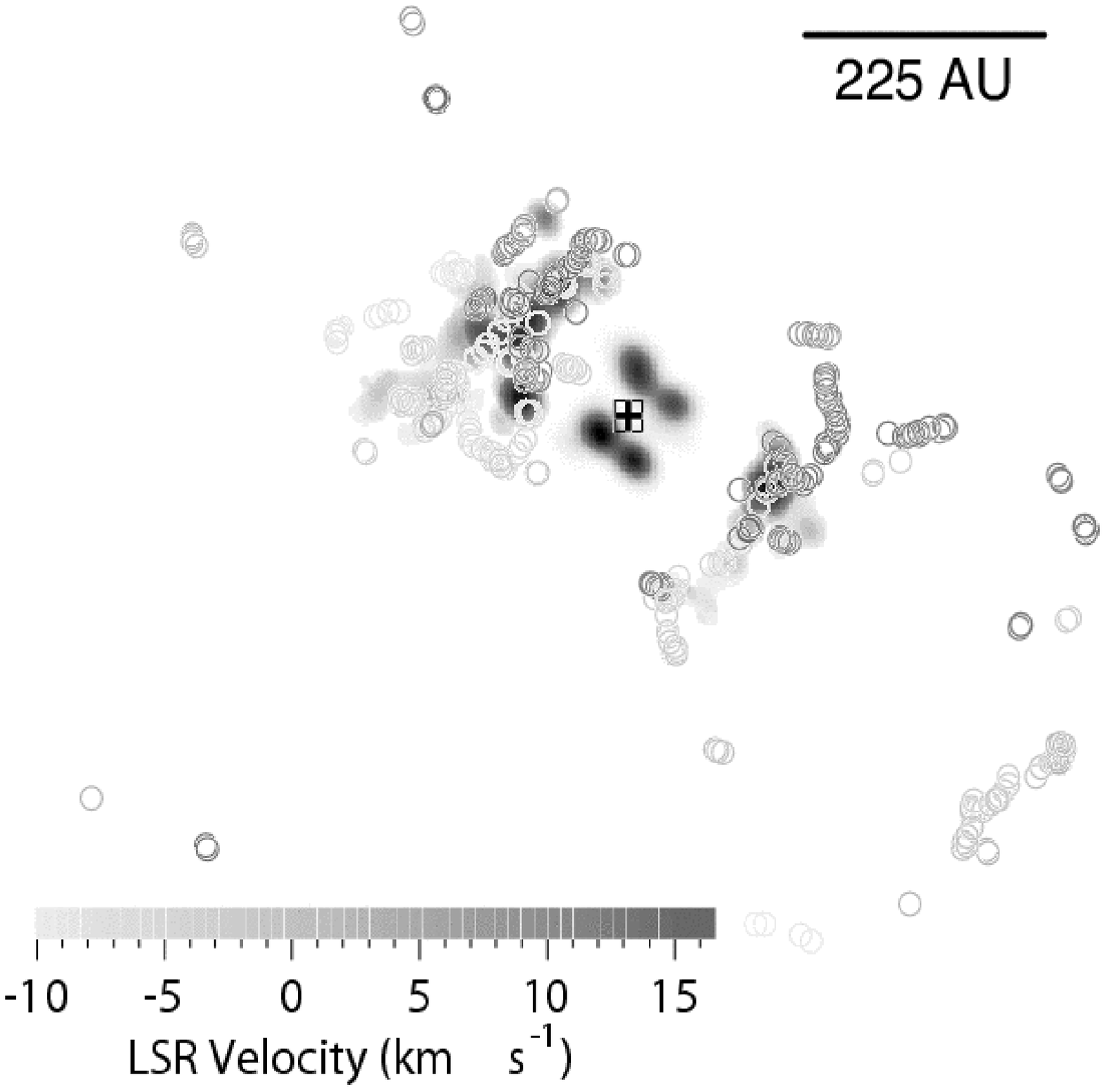}{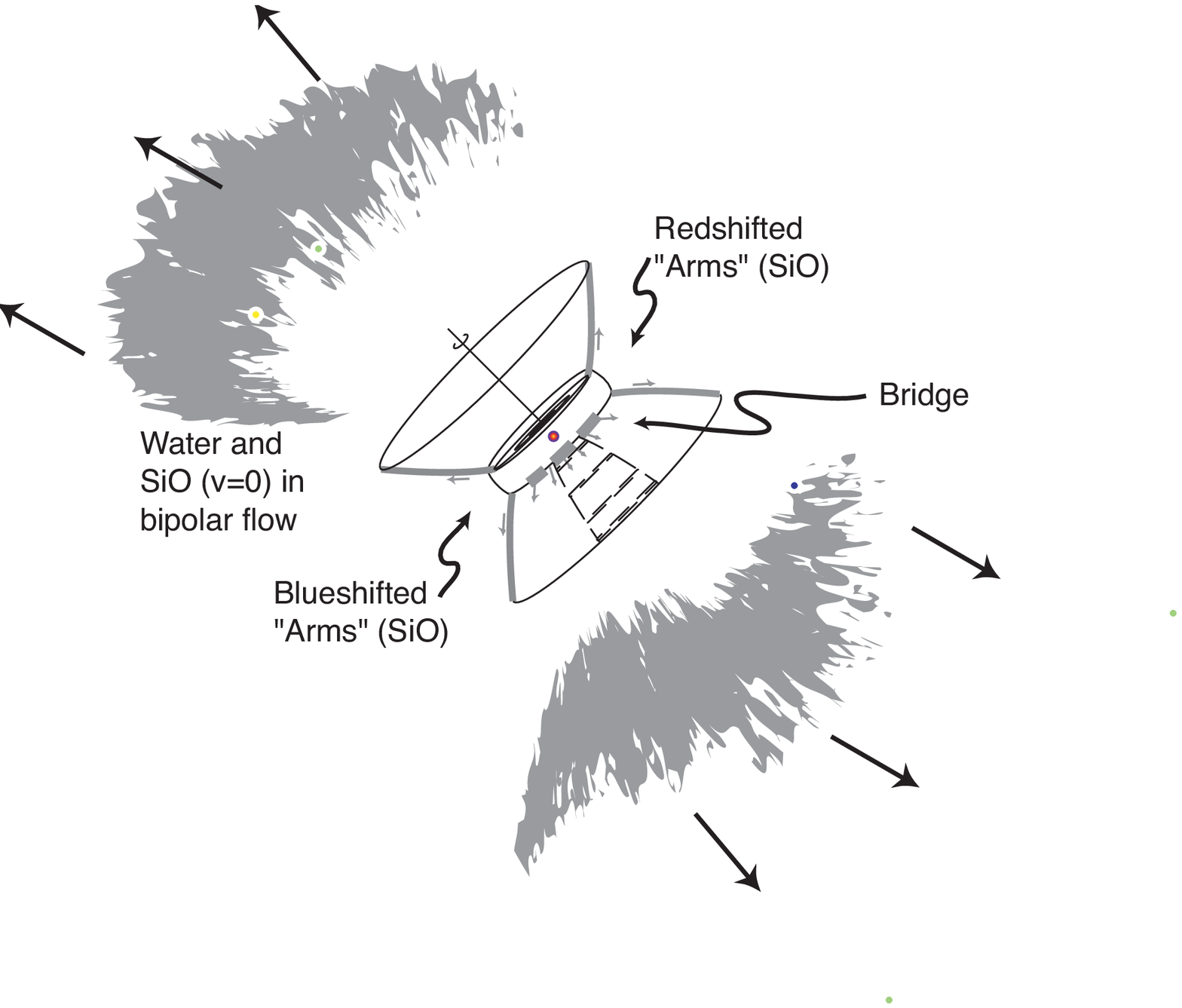}
\caption{ {\it (Left:)} Superposition of SiO maser emission integrated in velocity (grayscale)
and H$_2$O maser emission (circles). The ``cloverleaf'' near source~{\sf I} (the
box and cross) is v=1 \& 2 SiO emission. The arcs to the northeast and southwest are v=0 emission.
The H$_2$O spots mark the
emission centroids in the spectral channels.  Redshifted emission is
gray, blueshifted is black. {\it(Right:)} Sketch of the proposed edge-on disk model,
indicating the loci of SiO and H$_2$O maser emission.}
\end{figure}

\section {Summary}

We have fully resolved the structure and dynamics of material at radii of 20--500\,AU from
radio source~{\sf I} in the Orion KL region.  Our maps
provide the most detailed picture yet of molecular material so close to a massive YSO.   We
propose a new model for source~{\sf I} in which the accretion disk is 
edge-on and a wide-angle wind feeds
a funnel-like rotating bipolar outflow.  
Consequently, the
case of source~{\sf I} provides strong evidence that accretion in high-mass star formation
proceeds via orderly disk-mediated accretion as opposed to coalescence of low mass stars
(Bonnell et al. 1998).

\end{document}